\documentclass[aps,pra,10pt,amsmath,onecolumn,notitlepage,tightenlines,superscriptaddress,floatfix,eqsecnum,showpacs]{revtex4-1}

\usepackage{amssymb}
\usepackage{accents}
\usepackage{subdepth}

\usepackage{xcolor}
\definecolor{dark-red}{rgb}{0.4,0.15,0.15}
\definecolor{dark-blue}{rgb}{0.15,0.15,0.4}
\definecolor{medium-blue}{rgb}{0,0,0.5}
\usepackage{hyperref}
\usepackage[all]{hypcap}%fix problems with figures and tables of the 'hyperref' package
\usepackage{cleveref}% Package hperref works wrongly within the 'subequations' environment without this.
\hypersetup{
    colorlinks, linkcolor={dark-red},
    citecolor={dark-blue}, urlcolor={medium-blue}
} %These setups replace the ugly and distracting red and green boxes of the 'hyperref' package with colored texts.
\newcommand{\nsp}{\hspace{-0.4pt}}
\newcommand{\ssp}{\hspace{0.4pt}}

\newcommand{\ket}[1]{\lvert\ssp #1\ssp \rangle}
\newcommand{\bra}[1]{\langle\ssp #1\ssp \rvert}
\newcommand{\norm}[1]{\lvert #1 \rvert}

\newcommand{\commut}[2]{[\ssp #1\ssp,\,#2\ssp]}
\newcommand{\commutb}[2]{\big[\ssp #1\ssp,\,#2\ssp\big]}

\newcommand{\anticommut}[2]{\{ #1\ssp,\,#2\}}
\newcommand{\anticommutb}[2]{\big\{ #1\ssp,\,#2\big\}}

\newcommand{\identity}{\openone}
\newcommand{\half}{\frac{1}{2}}
\newcommand{\dt}[1]{\accentset{\vspace{0.5pt}\hspace{0.5pt}\mbox{\Large .}}{#1}} %

\newcommand{\apos}[1]{{#1'}} 
\newcommand{\thickbar}[1]{\accentset{\rule{.5em}{.6pt}}{#1}}

\newcommand{\sE}{\mathcal{E}}
\newcommand{\sR}{\mathcal{C}}
\newcommand{\sI}{\mathcal{I}}
\newcommand{\sL}{\mathcal{L}}
\newcommand{\dif}{d}
\newcommand{\idf}{f}
\newcommand{\idG}{G}
\newcommand{\idV}{V}
\newcommand{\idg}{g}
\newcommand{\idt}{t}
\newcommand{\xibf}{\boldsymbol\xi}
\newcommand{\std}{\mathrm{s}}

\DeclareMathOperator{\tr}{tr}
\DeclareMathOperator{\diag}{diag}

\begin{document}

\title{\bf Quantum Fisher information for states in exponential form}

\date{\today}

\author{Zhang Jiang}
\affiliation{Center for Quantum Information and Control, University of New Mexico, MSC07-4220, Albuquerque, New Mexico 87131-0001, USA}

\begin{abstract}
 We derive explicit expressions for the quantum Fisher information and the symmetric logarithmic derivative (SLD) of a quantum state in the exponential form $\rho = \exp (\idG)$; the SLD is expressed in terms of the generator $\idG$.  Applications include quantum-metrology problems with Gaussian states and general thermal states.  Specifically, we give the SLD for a Gaussian state in two forms, in terms of its generator and its moments; the Fisher information is also calculated for both forms.  Special cases are discussed, including pure, degenerate, and very noisy Gaussian states.
\end{abstract}

% Measurement theory (quantum mechanics), 03.65.Ta
% Quantum information, 03.67.-a
% Metrology, 06.20.-f
%% Interferometry, nonclassical, 42.50.St
%% Error theory, 06.20.Dk

\maketitle

\section{Introduction}
\label{sec:intro}

Quantum metrology studies the limit to the accuracy, set by quantum mechanics, with which physical quantities can be estimated by measurements.  The basic idea is to determine an unknown parameter~$\theta$ by probing a quantum state that depends on the parameter.  Quantum metrology is important for various purposes, which include improving time and frequency standards~\cite{udem_optical_2002, hinkley_atomic_2013}, detecting gravitational waves~\cite{ligo_enhanced_2013, andersen_quantum_2013}, interferometry based on interacting systems~\cite{dunningham_sub-shot-noise-limited_2004, riedel_atom-chip-based_2010}, and magnetometry~\cite{maze_nanoscale_2008, taylor_high-sensitivity_2008}.

A standard scenario for quantum parameter estimation is to put a known initial state $\rho_{\ssp \mathrm{in}}$ through a quantum channel $\sE_\theta$ that impresses $\theta$ on the system; the output state $\rho(\theta) = \sE_\theta(\rho_{\ssp \mathrm{in}})$ is then subjected to a measurement.  The goal is to find the optimal measurement strategy so that as much information as possible about $\theta$ is acquired.   Although it is hard to solve the most general problem exactly, bounds on how accurately one can estimate a parameter can be obtained~\cite{boixo_generalized_2007, tsang_fundamental_2011, escher_general_2011, giovannetti_quantum_2012}.

In classical parameter estimation theory, the Cram\'{e}r-Rao bound (CRB) expresses a lower bound on the variance of an unbiased estimator $\theta_\mathrm{est}$,
\begin{align}
 \mathrm{var}(\theta_\mathrm{est}) \geq \frac{1}{\sI_c(\theta)} \;,
\end{align}
where $\sI_c(\theta)$ is the classical Fisher information~\cite{fisher_information}.  Fisher's theory says that maximum likelihood estimation achieves the CRB asymptotically for large number of trials~\cite{fisher_theory_1925, cramer_mathematical_1999}.  For the quantum case, it was shown, in~\cite{braunstein_statistical_1994}, that there exists an optimal quantum measurement whose classical Fisher information, obtained from the measurement outcomes, achieves the quantum Fisher information~\cite{helstrom_quantum_1976, Holevo_probabilistic_1982, braunstein_statistical_1994, braunstein_generalized_1996},
\begin{align}\label{eq:Fisher}
 \sI(\theta) = \tr \big(\rho(\theta) L^2(\theta) \big)\;.
\end{align}
Thus the inverse of the quantum Fisher information gives the quantum CRB on the variance of an estimator.  The (Hermitian) operator $L(\theta)$, in Eq.~(\ref{eq:Fisher}), is the symmetric logarithmic derivative (SLD), defined implicitly by
\begin{align}\label{eq:SLD_def}
\frac{\dif \rho(\theta)}{\dif \theta} = \half\ssp \anticommutb{L(\theta)}{\rho(\theta)}\;,
\end{align}
where the brackets denote the anticommutator.  Knowing the SLD allows one to obtain not only the Fisher information but also the optimal measurement scheme.

Any full rank quantum state $\rho(\theta)$ can be written in exponential form,
\begin{align}\label{eq:exponential_rho}
 \rho(\theta) = e^{\idG(\theta)}\;,
\end{align}
with the normalization absorbed into $G(\theta)$.  The case that $\rho(\theta)$ is not invertible can be handled as a limit in which some eigenvalues of $G(\theta)$ go to minus infinity.  The form~(\ref{eq:exponential_rho}) is useful when $G(\theta)$ takes a simple form, examples being Gaussian states and general thermal states.  Gaussian states are important because of their appealing properties for quantum-metrology tasks~\cite{aspachs_phase_2009, anisimov_quantum_2010, lang_optimal_2013} and their accessibility both to experimentalists and theorists.  Thermal states are also useful for quantum-metrology tasks for at least two reasons: (i) The initial state is often a thermal state $\rho_{\ssp \mathrm{in}} = e^{-\beta H}/Z$, and the simple exponential form is preserved by a unitary channel $U_\theta$.  (ii) We can infer the temperature and the chemical potential by measuring the state $\rho(\theta) = e^{-\beta(H-\mu N)}/Z$, after the system is brought to thermodynamic equilibrium with a reservoir~\cite{Ugo_precision_2013}.

In Sec.~\ref{sec:exponential_density_matrix}, we consider the SLD for a quantum state in the exponential form~(\ref{eq:exponential_rho}).  We show that the SLD can be expanded into a weighted sum of $\dif \idG/\dif \theta$ and its recursive, nested commutators with $\idG$. Simple expressions of the quantum Fisher information and the SLD are given in the basis where $\idG$ is diagonalized.  In Sec.~\ref{sec:Gaussian_states_a}, we apply the results of Sec.~\ref{sec:exponential_density_matrix} to Gaussian states, and an explicit expression of the SLD in terms of the generator is derived.  In Sec.~\ref{sec:Gaussian_states_b}, also for Gaussian states, the SLD and the quantum Fisher information are given in terms of the moments of position and momentum operators (or of creation and annihilation operators).

\section{Quantum Fisher information for states in exponential form}
\label{sec:exponential_density_matrix}

A useful expression (see Eq.~(2.1) of Ref.~\cite{wilcox_exponential_1967}) for density operators of the exponential form~(\ref{eq:exponential_rho}) is
\begin{align}\label{eq:wilcox}
 \dt \rho = \int_0^1 e^{s\idG}\,\dt \idG\, e^{(1-s)\idG}\,\dif s\;,
\end{align}
where an overdot denotes a derivative with respect to $\theta$.  We now use the nested-commutator relation
\begin{align}
e^G A e^{-G}&=A+[G,A]+\frac{1}{2!}\,\big[G,[G,A]\ssp\big]+\cdots\nonumber\\
&=\sum_{n=0}^\infty\frac{1}{n!}\,\sR^n(A)=e^{\sR}(A)\;,
\end{align}
where $\sR^n(A)$, a linear operation on $A$, denotes the $n$th-order nested commutator $\commutb{\idG}{\nsp\ldots\ssp,\commut{\idG}{A}}$, with $\sR^0(A)=A$.  Applying this relation to the expression~(\ref{eq:wilcox}), we get
\begin{align}\label{eq:rhodot}
\begin{split}
 \dt \rho \rho^{-1}
 &= \dt \idG +\frac{1}{2!}\,\commut{\idG}{\dt \idG}+\frac{1}{3!}\,\commutb{\idG}{\commut{\idG}
 {\dt\idG}}+\cdots\\
 &= \sum_{n=0}^\infty \frac{1}{(n+1)!}\, \sR^n(\dt G)
 = h(\sR)(\dt\idG) \;,
\end{split}
\end{align}
where $h$ is the generating function of the expansion coefficients in Eq.~(\ref{eq:rhodot}),
\begin{align}\label{eq:generating_g}
 h(\idt) = 1+\frac{\idt}{2!}+\frac{\idt^2}{3!}+\cdots = \frac{e^\idt - 1}{\idt}\;.
\end{align}

Using the definitions~(\ref{eq:SLD_def}) and (\ref{eq:exponential_rho}), we also have
\begin{align}\label{eq:left_derivative_b}
\begin{split}
\dt \rho\ssp \rho^{-1}  &= \half\big(L + e^G L e^{-G}\big)\\[2pt]
&= \half \bigg(L + \sum_{n=0}^\infty \frac{1}{n!}\, \sR^n(L) \bigg)
= r(\sR)(L)\;,
\end{split}
\end{align}
where the generating function is $r(t)=(e^\idt+1)/2$.  Suppose that the SLD adopts the form,
\begin{align}\label{eq:expansion_L_a}
L &= \sum_{n=0}^\infty \idf_n\ssp \sR^n(\dt \idG)
=f(\sR)(\dt\idG)\;,
\end{align}
where the to be determined generating function $f$ is specified by
\begin{align}\label{eq:f_expansion_a}
 f(\idt) = \idf_0+\idf_1 \idt+ \idf_2 \idt^2+\cdots\;.
\end{align}

By putting Eq.~(\ref{eq:expansion_L_a}) into Eq.~(\ref{eq:left_derivative_b}), we have
\begin{align}\label{eq:combining_expansions}
 \dt \rho\ssp \rho^{-1}= r(\sR)\bigl[f(\sR)(\dt\idG)\bigr] =r\cdot f(\sR)(\dt\idG)\;,
\end{align}
where $r\cdot f$ is the product of the two functions, and we use the identity $\sR^n(\sR^m(A))=\sR^{n+m}(A)$. Comparing Eq.~(\ref{eq:combining_expansions}) with Eq.~(\ref{eq:rhodot}), we have the relation among the generating functions,
\begin{align}\label{eq:f_expansion_b}
 f(\idt)= \frac{h(t)}{r(t)}=\frac{\tanh(\idt/2)}{\idt/2}
 = \sum_{n=0}^\infty \frac{4\ssp (4^{n+1}-1) B_{2n+2}}{(2n+2)!}\,\idt^{2n}\;,
\end{align}
where $B_{2n+2}$ is the $(2n+2)$th Bernoulli number.  Comparing Eqs.~(\ref{eq:f_expansion_a}) with (\ref{eq:f_expansion_b}), we have
\begin{align}
 \idf_n =
\begin{cases}
 \displaystyle{\frac{4\ssp (4^{n/2+1}-1) B_{n+2} }{(n+2)!}}\,,&\mbox{for even $n$}\,,\\[3pt]
  0\,,&\mbox{for odd $n$}\,.
\end{cases}
\end{align}
The vanishing of the odd-order $\idf_n$s is a consequence of the Hermiticity of $L$, which makes $f(t)$ an even function.

The first four nonzero coefficients $\idf_n$ are
\begin{align}
 \idf_0 = 1\,,\quad  \idf_2=-\frac{1}{12}\,,\quad  \idf_4=\frac{1}{120}\,,\quad  \idf_6=-\frac{34}{8!}\;.
\end{align}
Although it appears that the $\idf_n$s become negligible very fast, they revive at larger $n$, and the radius of convergence of the power series~(\ref{eq:f_expansion_a}) is $t<\pi$.  This limits the usefulness of the expansion~(\ref{eq:expansion_L_a}); it is divergent when the difference between any two eigenvalues of $G$ is greater than or equal to $\pi$.  Fortunately, in many real problems, the recursive commutators in Eq.~(\ref{eq:expansion_L_a}) either terminate or repeat, enabling us to find an exact solution.  In the latter case, we can use analytic continuation to extend the result~(\ref{eq:expansion_L_a}) beyond the domain of convergence.

% From the definition of the SLD~(\ref{eq:SLD_def}), a changing of basis,
% \begin{align}\label{eq:basis_transformation_a}
%  \idG\rightarrow \idG' = U \idG U^\dagger\,,\quad \dt \idG\rightarrow \dt \idG' = U \dt \idG U^\dagger\;,
% \end{align}
% where $U$ is a unitary independent of $\theta$, leads to
% \begin{align}\label{eq:basis_transformation_b}
%  L\rightarrow L' = U L U^\dagger\;.
% \end{align}
% In the basis where $\idG$ is diagonalized, Eq.~(\ref{eq:expansion_L_a}) is equivalent to

Suppose that we work in the basis $\ket{e_j}$ where $\idG$ is diagonal, i.e., $\idG\ssp \ket{e_j}=g_j\ssp \ket{e_j}$.  This basis generally changes with $\theta$, so we are considering here, as in the rest of this section, a particular value of $\theta$.  In this basis, Eq.~(\ref{eq:expansion_L_a}) is equivalent to
\begin{align}\label{eq:expansion_L_b}
L_{jk}=\bra{e_j} L \ket{e_k} 
=f(\idg_j-\idg_k)\dt \idG_{jk}\;.
\end{align}
The domain of Eq.~(\ref{eq:expansion_L_b}) is not restricted to the radius of convergence, $\idg_j-\idg_k < \pi$; it is well defined for any $\idG$, which is an example of analytic continuation. Using Eq.~(\ref{eq:rhodot}), we have
\begin{align}
\dt\rho_{jk}=\bra{e_j}\dt\rho\ket{e_k}=e^{g_k}h(g_j-g_k)\ssp \dt\idG_{jk}
%=\frac{e^{g_j}-e^{g_k}}{g_j-g_k}\dt\idG_{jk}\;,
\end{align}
and Eq.~(\ref{eq:expansion_L_b}) can be converted to a formula familiar from Ref.~\cite{braunstein_statistical_1994},
\begin{align}
L_{jk} = \frac{\dt\rho_{jk}}{e^{g_k}r(g_j-g_k)}
=\frac{2\dt\rho_{jk}}{\rho_{jj}+\rho_{kk}}\;,
\end{align}
where $\rho_{jj} = e^{g_j}$.  This formula follows directly from the definition~(\ref{eq:SLD_def}) of the \hbox{SLD}.

The Fisher information can now be calculated directly in this same basis,
\begin{align}\label{eq:Fisher_information}
\begin{split}
 \sI
 &= \sum_{j,k} e^{\idg_j} \norm{L_{jk}}^2 = \sum_{j,k} e^{\idg_j}\ssp f^2(\idg_j-\idg_k)\ssp \norm{\dt \idG_{jk}}^2 \;.\\
 %= \sum_{j,k} \frac{2|\dt\rho_{jk}|^2}{\rho_{jj}+\rho_{kk}}
\end{split}
\end{align}

As a simple example, we discuss the SLD and Fisher information for a qubit.  Letting the Pauli matrices be denoted by $\sigma_j$, we can, without loss of generality, assume that the qubit state is diagonal in the eigenbasis of $\sigma_3$ and write the state as $\rho=\frac12(\sigma_0+\sigma_3\tanh\gamma)$, where $\tanh\gamma$ is the expectation value of $\sigma_3$.  This gives us
\begin{align}
 &\idG = \gamma \sigma_3 - \ln(2\cosh \gamma)\sigma_0\;,\label{eq:R_qubit}\\
 &\dt \idG = \dt \gamma \big(\sigma_3 - \sigma_0 \tanh \gamma \big)+ \tau_1\sigma_1 + \tau_2\sigma_2\;.\label{eq:dtR_qubit}
\end{align}
Here $\dt\gamma$ accounts for the change in the eigenvalues of $\rho$ as $\theta$ changes, and the real parameters $\tau_1$ and $\tau_2$ account for the change in eigenbasis of $\rho$ as $\theta$ changes.  Putting Eqs.~(\ref{eq:R_qubit}) and (\ref{eq:dtR_qubit}) into Eq.~(\ref{eq:expansion_L_a}) or (\ref{eq:expansion_L_b}), we have
\begin{align}
 L = \dt \gamma \big(\sigma_3 - \sigma_0 \tanh \gamma \big)+ \frac{\tanh \gamma}{\gamma}\ssp \big(\tau_1\sigma_1 + \tau_2\sigma_2\big)\;.
\end{align}
This expression can be verified by expanding the $2\times 2$ density operator explicitly.  The result for the Fisher information is
\begin{align}
\sI=\frac{\dt\gamma^2}{\cosh^2\!\gamma}+\frac{\tanh^2\!\gamma}{\gamma^2}(\tau_1^2+\tau_2^2)\;.
\end{align}

When the eigenvalues of the density operator $\rho$ are independent of $\theta$, i.e., the change of $\rho$ can be described by a unitary process, we have $\dt G = i\, \commut{G}{H}=i\, \mathcal{C}(H)$, where $H$ is some Hermitian operator. Putting this expression into Eq.~(\ref{eq:expansion_L_a}), we have the following formula for the SLD:
\begin{equation}
 L = f(\mathcal{C})(\dt G) = i\ssp f(\mathcal{C})\ssp \mathcal{C} (H)=2i \tanh (\mathcal{C}/2)(H)\;,\tag{2.20}
\end{equation}
which was first found by Knysh and Durkin (see Eq.~(A3) of Ref.~\cite{knysh_estimation_2013}).

\section{Gaussian states in exponential form}
\label{sec:Gaussian_states_a}

In this section, we apply the expansion~(\ref{eq:expansion_L_a}) to Gaussian states, which naturally adopt the exponential form,
\begin{align}\label{eq:Gaussian_exponential_a}
 \rho = e^G = \ssp \exp\Big(\mathord{-}\half\ssp \mathbf r^T\! \Omega\, \mathbf r + \mathbf r^T \boldsymbol \eta-\ln Z\Big)\;,
\end{align}
where $\mathbf r =\big(x_1\;\cdots\; x_n\:\, p_1\;\cdots\; p_n\big)^T$ is the $2n$-dimensional vector of position and momentum operators, $\boldsymbol \eta$ is a real $2n$-dimensional vector, and $\Omega> 0$ is a $2n\times 2n$ real, symmetric matrix.  The state~(\ref{eq:Gaussian_exponential_a}) can be regarded as a thermal state, with $\beta=1$, of the quadratic Hamiltonian
\begin{align}\label{eq:Gaussian_Hamiltonian_a}
H=\half\ssp \mathbf r^T\! \Omega\, \mathbf r - \mathbf r^T \boldsymbol \eta\;;
\end{align}
notice that $Z=\tr (e^{-H}\ssp)$.

The canonical commutation relations can be written as $\commut{r_j}{r_k} = i J_{jk}$, where $J$ is the skew-symmetric matrix
\begin{align}
 J =
\begin{pmatrix}
  0          & \identity\\
  -\identity & 0
 \end{pmatrix}
=-J^T=-J^{-1}\;,
\end{align}
with $\identity$ being the $n\times n$ identity matrix.  For any Gaussian state, both $\idG$ and $\dt \idG$ are degree-$2$ polynomials of the position and momentum operators, and thus so are all the recursive commutators in Eq.~(\ref{eq:expansion_L_a}).  Consequently, $L$ is also a degree-$2$ polynomial of the position and momentum operators,
\begin{align}\label{eq:L_Gaussian}
 L = \mathbf r^T\nsp \Phi \, \mathbf r\ + \mathbf r^T\! \boldsymbol \zeta -\nu \;,
\end{align}
where $\boldsymbol \zeta$ is a real $2n$-dimensional vector, and $\Phi$  is a $2n\times 2n$ real, symmetric matrix, and $\nu$ can be determined by the trace-preserving condition,
\begin{align}
 \nu = \tr\big(\rho\, \mathbf r^T\nsp \Phi \, \mathbf r\big)\;.
\end{align}

In order to use the expansion~(\ref{eq:expansion_L_a}) efficiently, we write the quadratic Hamiltonian in the basis of creation and annihilation operators,
\begin{align}
H=\half\ssp\ssp \thickbar{\boldsymbol a}\, \apos\Omega\ssp \boldsymbol a - \thickbar{\boldsymbol a}\,\apos{\boldsymbol \eta}\;,\label{eq:Gaussian_Hamiltonian_b}
\end{align}
where $\boldsymbol a$ and $\thickbar{\boldsymbol a}$ are vectors of the creation and annihilation operators,
\begin{align}
&\thickbar{\boldsymbol a} = (a_1^\dagger\,\cdots\, a_n^\dagger\: a_1\,\cdots\, a_n)\;,\\
&\boldsymbol a = (a_1\,\cdots\, a_n\: a_1^\dagger\,\cdots\, a_n^\dagger)^T\;,
\end{align}
with $a_j=(x_j+ip_j)/\sqrt 2\,$; the matrix $\apos\Omega$ and the vector $\apos{\boldsymbol \eta}$ satisfy
\begin{align}
\apos\Omega = \idV^\dagger \Omega\ssp \idV\,,\quad \apos{\boldsymbol \eta} = \idV^\dagger \boldsymbol \eta\;,
\end{align}
where $\idV$ is a unitary matrix linking the two bases, i.e., $\idV \ssp \boldsymbol a = \mathbf r$, or equivalently, $\idV^\dagger \mathbf r = \boldsymbol a$,
\begin{align}
 \idV^\dagger =  \frac{1}{\sqrt 2}\,
 \begin{pmatrix}
  \identity & \;i\identity\\
  \identity & \mathord{-}i \identity
 \end{pmatrix}\;.
\end{align}
Similarly, we can write the SLD as
\begin{align}
L = \thickbar{\boldsymbol a}\, \apos\Phi \boldsymbol a + \thickbar{\boldsymbol a}\, \apos{\boldsymbol \zeta}-\nu\;,\label{eq:L_Gaussian_b}
\end{align}
where $\apos \Phi = V^\dagger \Phi V$, and $\apos{\boldsymbol \zeta} = V^\dagger \boldsymbol \zeta$.

Without affecting the Fisher information, which is invariant under unitary transformations, we can displace the state~(\ref{eq:Gaussian_exponential_a}) so that $\boldsymbol\eta=0$.  Moreover, we now assume that the matrix $\Omega$ is in the diagonal form,
\begin{align}\label{eq:standard_Omega}
 \Omega  =
 \begin{pmatrix}
  \diag(\epsilon_1,\ldots, \epsilon_n)& 0\\[2pt]
  0 & \diag(\epsilon_1,\ldots, \epsilon_n)
 \end{pmatrix}= \apos\Omega\;,
\end{align}
which gives
\begin{align}\label{eq:diagonalized_form}
 \idG = -H-\ln Z
 = -\sum_{j = 1}^n \epsilon_j \Big(a_j^\dagger a_j + \half\, \Big) - \ln Z\;.
\end{align}
This case is important, because any Gaussian state is equivalent to it up to a Gaussian unitary, i.e., a symplectic transformation of the creation and annihilation operators.  The commutation relations between $\idG$ and the creation and annihilation operators are straightforward:
\begin{align}\label{eq:commutation_relations}
 \commutb{\idG}{a_j} = \epsilon_j a_j\,,\quad \commutb{\idG}{a_j^\dagger} = - \ssp \epsilon_j a_j^\dagger\;.
\end{align}
Consequently, we have
\begin{align}
 f(\mathcal C)\ssp (a_j) = f(\epsilon_j)\ssp a_j\;,\quad
 f(\mathcal C)\ssp (a_j^\dagger) %= f(-\epsilon_j)\ssp a_j^\dagger 
 = f(\epsilon_j)\ssp a_j^\dagger\;,\label{eq:relation_a}
\end{align}
and for quadratic operators, we have
\begin{align}
 f(\mathcal C)\ssp(a_j^\dagger a_k) = f(\epsilon_k-\epsilon_j)\ssp a_j^\dagger a_k\;,\\[2pt]
 f(\mathcal C)\ssp(a_j a_k) = f(\epsilon_j+\epsilon_k)\ssp a_j a_k\;.\label{eq:relation_b}
\end{align}
Most generally, the derivative of $G$ takes the form
\begin{align}\label{eq:derivative_G_a}
 \dt \idG=-\half\ssp\ssp \thickbar{\boldsymbol a}\, \apos{\dt\Omega}\ssp \boldsymbol a +\thickbar{\boldsymbol a}\, \apos{\dt{\boldsymbol \eta}}-\frac{\dt Z}{Z}\;,
\end{align}
Putting Eq.~({\ref{eq:derivative_G_a}) into Eq.~(\ref{eq:expansion_L_a}) and using the relations~(\ref{eq:relation_a})--(\ref{eq:relation_b}), we have
\begin{align}\label{eq:SLD_Gaussian_diagonal_a}
\nu = \dt Z/Z\,,\quad \apos{\boldsymbol \zeta_j}  = f(\epsilon_j)\ssp\ssp \apos{\dt{\boldsymbol \eta}_j}\;,
\end{align}
 and
\begin{align}\label{eq:SLD_Gaussian_diagonal_b}
\apos{\Phi_{jk}}  =
 \begin{cases}
-\half\ssp f(\epsilon_j-\epsilon_k) \ssp\ssp \apos{\dt{\Omega}_{jk}}\,,&\mbox{for $j,k\leq n$ or $j,k> n$}\,,\\[4pt]
-\half\ssp f(\epsilon_j+\epsilon_k) \ssp\ssp \apos{\dt{\Omega}_{jk}}\,,&\mbox{for all other cases}\,,
\end{cases}
\end{align}
where $\epsilon_{j+n} = \epsilon_j$ for $j \leq n$.  Equations~(\ref{eq:SLD_Gaussian_diagonal_a}) and  (\ref{eq:SLD_Gaussian_diagonal_b}) are explicit, and the only work required is to find the basis of the creation and annihilation operators, by a symplectic transformation, so that the Gaussian state is of the diagonal form~(\ref{eq:diagonalized_form}).

Knowing the SLD allows one to calculate the Fisher information [see Eq.~(\ref{eq:Fisher_Gaussian_b})],
\begin{align}
\sI &= \half \tr( \Gamma \Phi \Gamma \Phi)  -  \half ( J \Phi J^T\! \Phi) + \half\, \boldsymbol \zeta^T \Gamma \boldsymbol\zeta\;,
\end{align}
where $\Gamma$ is the covariance matrix of the Gaussian state defined in Eq.~(\ref{eq:variance}).  Going to the basis of creation and annihilation operators, we have
\begin{align}
\sI &= \half \tr( \apos \Gamma \apos  \Phi \apos \Gamma \apos \Phi)   -  \half (  \apos J \apos \Phi  \apos J \apos \Phi) + \half\, \apos {\boldsymbol \zeta}^\dagger \apos \Gamma\ssp \apos {\boldsymbol\zeta}\;,
\end{align}
where
\begin{align}
 \apos J = \apos J^{\, \dagger} =
\begin{pmatrix}
  \identity      &    0\\
     0             &   -\identity
 \end{pmatrix}\;,
\end{align}
and for $\apos \Omega$ taking the form~(\ref{eq:standard_Omega}), we have $\apos \Gamma = V^\dagger \Gamma V = \coth(\apos \Omega/2)$.  Thus, the Fisher information can be calculated explicitly,
\begin{widetext}
\begin{align}\label{eq:fisher_gaussian_generator}
 \sI
 &= \sum_{j,k=1}^n \big(\norm{\apos{\Phi_{jk}}}^2+\norm{\apos{\Phi_{j,k+n}}}^2\big) \coth\frac{\epsilon_j}{2}\coth\frac{\epsilon_k}{2}+\norm{\apos{\Phi_{j,k+n}}}^2-\norm{\apos{\Phi_{jk}}}^2 + \sum_{j=1}^n\, \norm{\apos{\boldsymbol\zeta_j}}^2 \coth\frac{\epsilon_j}{2}\;.
\end{align}
\end{widetext}
%
%\begin{widetext}
%\begin{align}
% \sI
% &= \sum_{j=1}^n\, \norm{\beta_j'}^2 \coth\frac{\epsilon_j}{2}+\sum_{j,k=1}^n \frac{1}{4}\,\big(\norm{\phi_{jk}'}^2+\norm{\varphi_{jk}'}^2\big) \coth\frac{\epsilon_j}{2}\coth\frac{\epsilon_k}{2}+\frac{1}{4}\,\big(\norm{\varphi_{jk}'}^2-\norm{\phi_{jk}'}^2\big)\;.
%\end{align}
%\end{widetext}

\section{Gaussian states by moments}
\label{sec:Gaussian_states_b}

A number of authors have already discussed SLDs and quantum Fisher information for Gaussian states.  Monras and Paris~\cite{monras_optimal_2007} investigated the problem of loss estimation with displaced squeezed thermal states.  Pinel {\em et al.}~\cite{pinel_ultimate_2012, pinel_quantum_2013} discussed parameter estimation with pure Gaussian states of arbitrarily many modes and general single-mode Gaussian states.  Recently, Monras~\cite{monras_phase_2013} found an equation---in terms of the moments---for the SLD of the most general Gaussian state.  The Fisher information can be calculated once the SLD is known.  Here we confirm Monras' results by using a different, somewhat simpler approach.  Furthermore, we solve the resultant equation of the SLD with a symplectic transformation.  Special cases are also discussed, which include pure, degenerate, and very noisy Gaussian states.

Most generally, the symmetrically ordered characteristic function of a Gaussian quantum state takes the form
\begin{align}\label{eq:characteristic}
 \chi_S(\xibf) \equiv \tr\big(\rho\, e^{\ssp i\mathbf r^T\nsp \xibf}\ssp\big) = \exp\Big(\mathord{-}\frac{1}{4}\ssp \xibf^T \Gamma\ssp \xibf + i\boldsymbol \delta^T\nsp \xibf\Big)
\end{align}
where $\boldsymbol \delta$ is a real $2N$-dimensional vector, and $\Gamma > 0$ is a $2N\times 2N$ real, symmetric matrix.  The vector $\boldsymbol \delta$ and the matrix $\Gamma$ represent the means and the covariance matrix of the Gaussian state,
\begin{align}
 &\delta_j = \tr (\rho\ssp\ssp r_j)\;,\label{eq:mean}\\[2pt]
 &\Gamma_{jk} = \tr\big(\rho\, \anticommut{\Delta r_j}{\Delta r_k}\big)\label{eq:variance}\;,
\end{align}
where $\Delta r_j = r_j-\delta_j$.  Without loss of generality, the mean $\boldsymbol \delta$ can be removed by a displacement,
\begin{align}
& \rho\rightarrow  e^{-i\mathbf r^T\!\nsp J \boldsymbol \delta} \rho\, e^{i\mathbf r^T\!\nsp J \boldsymbol \delta}\;,
\end{align}
and we assume $\boldsymbol \delta = 0$ from now on.

\subsection{Calculating the SLD}

Taking a derivative with respect to $\theta$ on both sides of Eqs.~(\ref{eq:mean}) and (\ref{eq:variance}) and using the definition~(\ref{eq:SLD_def}), we have
\begin{align}
 &\dt\delta_j = \half\ssp \tr \big(\anticommut{\rho}{L}\, r_j\big) = \half\ssp \tr \big(\rho\, \anticommut{L}{r_j}\big)\;,\label{eq:derivative_a}\\[3pt]
 &\dt\Gamma_{jk} = \half\tr\Big(\anticommut{\rho}{L}\ssp \anticommut{r_j}{r_k}\Big) = \half\tr\Big(\rho\, \anticommutb{L}{\anticommut{r_j}{r_k}}\Big)\;.\label{eq:derivative_b}
\end{align}
To calculate the traces in Eqs.~(\ref{eq:derivative_a}) and (\ref{eq:derivative_b}), we introduce the following function, which we call the {\em partially symmetrically ordered characteristic function\/},
\begin{align}
\begin{split}\label{eq:symmetric_chi}
 \chi_P(\xibf_1,\xibf_2) &\equiv \half\ssp \tr\Big(\rho\,\anticommutb{e^{\ssp i\mathbf r^T\nsp \xibf_1}}{e^{\ssp i\mathbf r^T\nsp \xibf_2}}\Big)\\
 &= \chi_S(\xibf_1+\xibf_2)\,\cos\Big(\frac{1}{2}\ssp \xibf_1^T\nsp J\ssp \xibf_2\Big)\;.
\end{split}
\end{align}
Denoting the partial derivative with respect to the $j$th element of $\xibf_{1,2}$ by $\partial^{(1,2)}_j$ we have
\begin{align}
 &\dt \delta_j = -i\sL^{(1)}\ssp \partial^{(2)}_j\ssp \chi_P \big\vert_{\xibf_1 = \xibf_2 =0}\;,\label{eq:derivative_c}\\[3pt]
 &\dt \Gamma_{jk} = -2\ssp \sL^{(1)}\ssp \partial^{\ssp (2)}_{jk}\ssp \chi_P \big\vert_{\xibf_1 = \xibf_2 =0}\;,\label{eq:derivative_d}
\end{align}
where $\partial_{jk} = \partial_j \partial_k$ and
\begin{align}\label{eq:SLD_form}
 \sL = \mathord{-}\sum_{m,n} \Phi_ {mn}\, \partial_{mn} -i\sum_l \zeta_l\ssp \partial_l - \nu\;.
\end{align}
Putting Eq.~(\ref{eq:SLD_form}) into Eqs.~(\ref{eq:derivative_c}) and (\ref{eq:derivative_d}), we have
\begin{align}
 & \dt \delta_j = -\Big(\sum_l \zeta_l\ssp \partial^{(1)}_l\Big) \partial^{(2)}_j\chi_P \Big\vert_{\xibf_{1,2}=0}= \half \sum_l \Gamma_{jl}\, \zeta_l\;,\label{eq:derivative_e}\\[3pt]
 \begin{split}
 &\dt \Gamma_{jk} = 2\Big(\sum_{m,n} \Phi_ {mn}\, \partial^{(1)}_{mn}+\nu\Big)\ssp \partial^{\ssp (2)}_{jk} \ssp \chi_P \Big\vert_{\xibf_{1,2}=0}\\
 & \hspace{1.5em} = \Big(\half\ssp\tr(\ssp \Gamma\ssp \Phi)-\nu\Big)\ssp \Gamma_{jk} + (\ssp \Gamma\ssp \Phi\ssp \Gamma +  J\ssp \Phi\ssp J\ssp)_{jk}\;,
 \end{split}\label{eq:derivative_f}
\end{align}
where all the odd-order derivatives are neglected, because they vanish at $\xibf_{1}=\xibf_{2}=0$ for $\boldsymbol \delta = 0$.  By using the trace-preserving condition,
\begin{align}\label{eq:trace_preserving}
 0 = \tr(L \rho) = \sL \ssp \ssp \chi_S\ssp \big\vert_{\xibf=0} = \half \tr(\ssp\Gamma\ssp \Phi) - \nu\;,
\end{align}
we have the following matrix forms
\begin{align}
 &\dt {\boldsymbol \delta} = \half\ssp \Gamma \boldsymbol \zeta\;,\label{eq:derivative_g}\\
 &\dt \Gamma = \Gamma\ssp \Phi\ssp \Gamma -  J\ssp \Phi\ssp J^T \;,\label{eq:derivative_h}
\end{align}
Equation~(\ref{eq:derivative_h}) is an implicit matrix equation, which is generally hard to solve.  A way to circumvent such difficulty is by using a symplectic transformation.   Any covariance matrix $\Gamma$ can be brought into the following standard (canonical) form by a symplectic transformation $S$ satisfying $SJS^T= J$,
\begin{align}
 S\ssp \Gamma S^T = \Gamma_\std =
 \begin{pmatrix}
  \Lambda & 0\\
  0       & \Lambda
 \end{pmatrix}\;,
\end{align}
where $\Lambda = \diag(\lambda_1,\,\lambda_2,\ldots,\lambda_n) \geq \identity$ is a diagonal matrix (equality holds, i.e., $\lambda_j=1$ for $j=1,\ldots,n$, only for pure states).  In the basis that $\Gamma$ is standard, Eq.~(\ref{eq:derivative_h}) reads
\begin{align}
 \dt \Gamma_\std = \Gamma_\std \ssp \Phi_\std \ssp \Gamma_\std  -  J\ssp \Phi_\std \ssp J^T \;,\label{eq:derivative_i}
\end{align}
where $\dt \Gamma_\std =  S\ssp \dt \Gamma S^T$, and $J \Phi_\std J^T=  SJ\ssp \Phi J^T\! S^T$.  Noticing that $\Gamma_\std$ and $J$ commute, we have
\begin{align}
  \Gamma_\std  \dt \Gamma_\std \Gamma_\std + J\ssp \dt \Gamma_\std J^T = \Gamma_\std^2 \Phi_\std  \Gamma_\std^2-  \Phi_\std\;,
\end{align}
which can be solved explicitly since $\Gamma_\std$ is diagonal,
\begin{align}\label{eq:derivative_j}
(\Phi_\std)_{jk} = \frac{\big(\Gamma_\std \dt \Gamma_\std  \Gamma_\std + J\ssp \dt \Gamma_\std J^T\big)_{jk} }{\lambda_j^2\lambda_k^2-1}\;,
\end{align}
where $\lambda_{j+n}=\lambda_{j}$ for $j\leq n$.
%Another way of writing Eq.~(\ref{eq:derivative_h})---which is manifestly symplectic covariant---is
%\begin{align}
% \dt \Gamma  = \Gamma\ssp J^T\tilde \Phi J\ssp \Gamma -  \tilde \Phi\;,\label{eq:derivative_i}
%\end{align}
%where $\tilde \Phi = J\ssp \Phi\ssp J^T$; for any symplectic transformation $S$ satisfying $SJS^T= J$, we have
%\begin{align}\label{eq:covariance}
% \dt \Gamma' = S\ssp \dt \Gamma S^T = \Gamma' J^T \tilde \Phi' J \Gamma' -  \tilde \Phi'\;,
%\end{align}
%where $\tilde \Phi' = S\ssp \tilde \Phi\ssp S^T$, and $\Gamma' = S\ssp \Gamma\ssp S^T$.
%In the symplectic basis where $\Gamma$ is standard, we have
%\begin{align}
% \tilde \Phi_ \std = S\ssp \tilde \Phi\ssp S^T\,,\quad \dt \Gamma_\std = S \dt\Gamma S^T\;.
%\end{align}
Once $\Phi_ \std$ is determined in terms of $\Gamma_\std$ and $\dt \Gamma_\std$, an inverse symplectic transformation can transform it back to $\Phi$. To end this subsection, we discuss some special cases where Eq.~(\ref{eq:derivative_j}) can be simplified to forms which are manifestly symplectic covariant; this allows us to solve the SLD and the Fisher information without going to the standard basis.
%With Eq.~(\ref{eq:derivative_i}), we have
%\begin{align}\label{eq:SLD_standard_basis}
%(\tilde \Phi_\std)_{jk} = \frac{\big(\Gamma_\std J^T \dt \Gamma_\std J\, \Gamma_\std + \dt \Gamma_\std\big)_{jk} }{\lambda_j^2\lambda_k^2-1}\;.
%\end{align}

For a very noisy Gaussian state where all $\lambda_j\gg 1$, we have the following relations
\begin{align}
 \Phi_\std \approx \Gamma_\std^{-1} \dt \Gamma_\std \Gamma_\std^{-1} \;,
\end{align}
which is symplectic covariant and can be generalized to 
\begin{align}\label{eq:noisy_Gaussian_relation}
 \Phi \approx \Gamma^{-1} \dt \Gamma \Gamma^{-1}\;.
\end{align}

For the degenerate case where $\lambda_j = \lambda$ for all $j$, we have
%driven by a Gaussian unitary,
%\begin{align}\label{eq:degenerate_Gaussian_relation}
% \Phi_\std = \frac{\dt \Gamma_\std }{\lambda^2+1}\;.
%\end{align}
\begin{align}
\Phi_\std = \frac{1}{\lambda^4-1}\,\big(\lambda^2\ssp \dt \Gamma_\std  + J\ssp\ssp \dt \Gamma_\std J^T\ssp\big)\;,
\end{align}
which can be brought into the following symplectic covariant form,
\begin{align}\label{eq:derivative_k}
\Phi = \frac{1}{\lambda^4-1}\,\big(\lambda^4\ssp \Gamma^{-1}\dt \Gamma \Gamma^{-1}  + J\ssp\ssp \dt \Gamma J^T\ssp\big)\;,
\end{align}
If the symplectic eigenvalues of $\Gamma$ do not change, i.e., $\dt \Gamma$ is driven by some Gaussian unitary, we have
\begin{align}\label{eq:Gaussian_unitary}
 \dt \Gamma = \Gamma H J^T + J H \Gamma\;,
\end{align}
where $H=H^T$; this equation can be derived by considering the evolution of the covariance matrix~(\ref{eq:variance}) under the quadratic Hamiltonian $\mathbf r^T\! H \mathbf r/2$.  With  the condition $\lambda^2\ssp \Gamma^{-1} = J\ssp \Gamma J^T$ for degenerate Gaussian states and Eq.~(\ref{eq:Gaussian_unitary}), we have 
\begin{align}\label{eq:unitary_condition}
 \lambda^2\ssp \Gamma^{-1}\dt \Gamma \Gamma^{-1} = J \Gamma H  +  H\ssp \Gamma J^T = - J\ssp \dt \Gamma J^T\;,
\end{align}
and thus Eq.~(\ref{eq:derivative_k}) can be simplified to
\begin{align}\label{eq:derivative_l}
\Phi  = \frac{\lambda^2}{\lambda^2 + 1}\, \Gamma^{-1}\dt \Gamma \Gamma^{-1} = -\frac{1}{\lambda^2 + 1}\, J\ssp \dt \Gamma J^T\;.
\end{align}
%A special case is the pure state case where $\Gamma_\std = I \equiv  \identity \oplus \identity $, and Eq.~(\ref{eq:derivative_l}) is singular; this can be overcome by setting $L = 0$ in the nullspace of $\rho$, i.e., $\dt \Gamma$ is driven by some Gaussian unitary,

For pure Gaussian states ($\lambda = 1$), we assume that the condition Eq.~(\ref{eq:unitary_condition}) is always satisfied; otherwise, $\Phi$ would diverge according to Eq.~(\ref{eq:derivative_k}). By setting $\lambda =1$ in Eq.~(\ref{eq:derivative_l}), we have the following result for pure states:
\begin{align}\label{eq:pure_Gaussian_relation}
 \Phi  = \half\, \Gamma^{-1}\dt \Gamma \Gamma^{-1} = -\half\, J\ssp \dt \Gamma J^T\;.
\end{align}
%which can be derived by setting $\lambda = 1+\epsilon$ and taking the limit $\epsilon \rightarrow 0$.
Note that Eq.~(\ref{eq:pure_Gaussian_relation}) is valid even if the pure Gaussian state actually goes through a nonunitary process which gives the same $\dt \rho$ as a unitary process for that pure state.

\subsection{Quantum Fisher information}

The Fisher information can be calculated by applying $\sL$ on $\chi_P$ twice,
\begin{align}\label{eq:Fisher_Gaussian_a}
 \sI &= \tr(\rho L^2)= \sL^{(1)}\sL^{(2)} \ssp \chi_P \big\vert_{\xibf_{1,2}=0}\;.
\end{align}
Putting Eq.~(\ref{eq:SLD_form}) into Eq.~(\ref{eq:Fisher_Gaussian_a}) and neglecting all derivatives of odd orders, we have
\begin{widetext}
\begin{align}\label{eq:Fisher_Gaussian_b}
\begin{split}
\sI &= \Big(\sum_{j,k,l,m} \Phi_ {jk} \Phi_ {lm} \partial^{(1)}_{jk} \partial^{(2)}_{lm} + \sum_{j,k} 2\nu\ssp \Phi_ {jk} \partial^{(1)}_{jk}-\sum_{j,k}\zeta_j\zeta_k \partial^{(1)}_{j}\partial^{(2)}_{k}+\nu^2\Big)\ssp \chi_P \Big\vert_{\xibf_{1,2}=0}\\
&= \half \tr(\ssp J\Phi J\Phi)+\half \tr(\Gamma\ssp \Phi\ssp \Gamma\ssp \Phi) +\frac{1}{4} \big(\!\tr(\ssp \Gamma\ssp \Phi)\big)^2 -\nu \tr(\ssp \Gamma\ssp \Phi)+\half\ssp \boldsymbol \zeta^T \Gamma \boldsymbol\zeta+\nu^2\\[3pt]
&= \half \tr\Big(\big(\ssp \Gamma\ssp \Phi\ssp \Gamma-J\Phi J^T\big) \Phi\Big)+\half\ssp \boldsymbol \zeta^T \Gamma \boldsymbol\zeta\\[3pt]
&= \half \tr(\ssp \dt\Gamma\ssp \Phi)+2\ssp \dt{\boldsymbol \delta}^T \Gamma^{-1} \dt{\boldsymbol \delta}\;,
\end{split}
\end{align}
\end{widetext}
where the conditions~(\ref{eq:trace_preserving}), (\ref{eq:derivative_g}), and (\ref{eq:derivative_h}) are used to simplify the expressions; also note that the quantity $\tr(\ssp \dt\Gamma\ssp \Phi)$ is symplectically invariant, specifically,
\begin{align}\label{eq:Fisher_symplectic_invariant}
 \tr(\ssp \dt\Gamma\ssp \Phi) &= \tr(\ssp \dt\Gamma_\std\ssp  \Phi_\std )\;.
\end{align}

For very noisy Gaussian states, we have $\dt\Gamma \Phi = (\ssp \dt\Gamma \Gamma^{-1})^2$ by Eq.~(\ref{eq:noisy_Gaussian_relation}), and consequently, the quantum Fisher information reads
\begin{align}
 \sI_\mathrm{noisy} \approx \half \tr\big((\ssp \dt\Gamma \Gamma^{-1})^2\big)+2\ssp \dt{\boldsymbol \delta}^T \Gamma^{-1} \dt{\boldsymbol \delta}\;.
\end{align}  

For a degenerate Gaussian state, the quantum Fisher information can be derived by using Eq.~(\ref{eq:derivative_k}),
\begin{align}\label{eq:degenerate_general}
 \sI_\mathrm{degen} = \frac{\tr\big(\lambda^4(\ssp \dt\Gamma \Gamma^{-1})^2-(\dt\Gamma J)^2\big)}{2\ssp (\lambda^4-1)} +2\ssp \dt{\boldsymbol \delta}^T \Gamma^{-1} \dt{\boldsymbol \delta}\;.
\end{align}
If the degenerate Gaussian state is driven by a Gaussian unitary, we have
\begin{align}\label{eq:degenerate_a}
 \sI_\mathrm{degen} &= \frac{\lambda^2}{2\ssp (\lambda^2+1)} \tr\big((\ssp \dt\Gamma \Gamma^{-1})^2\big)+2\ssp \dt{\boldsymbol \delta}^T \Gamma^{-1} \dt{\boldsymbol \delta}\;.
\end{align}
or equivalently,
\begin{align}\label{eq:degenerate_b}
 \sI_\mathrm{degen} &= \frac{1}{2\ssp (\lambda^2+1)}\,\tr\big((\ssp \dt\Gamma J)^2\big) +\frac{2}{\lambda^2}\, \dt{\boldsymbol \delta}^T\! J\ssp \Gamma J^T \dt{\boldsymbol \delta}\;,
\end{align}
where we use the identity $\Gamma^{-1} = J\ssp \Gamma J^T/\lambda^2$ for degenerate Gaussian states. In particular, Eqs.~(\ref{eq:degenerate_a}) and (\ref{eq:degenerate_b}) work for all single-mode Gaussian states.  

For pure Gaussian states, we have
%the relation~(\ref{eq:pure_Gaussian_relation}), and consequently,
%\begin{align}
% \tr(\ssp \dt\Gamma\ssp \Phi)  = \half\ssp \tr(\ssp \dt\Gamma_\std^2)\;.
%\end{align}
%The covariance matrix of any pure Gaussian state is symplectic equivalent to the identity; $S \Gamma S^T = I$, or equivalently, $\Gamma^{-1} = S^T\! S$, and we have
%\begin{align}
% \tr(\ssp \dt\Gamma_\std^2) &= \tr(\ssp S\ssp \dt\Gamma S^T\! S\ssp \dt\Gamma S^T) = \tr\big((\ssp \dt\Gamma \Gamma^{-1})^2\big)\;.
%\end{align}
%Thus, the Fisher information for pure Gaussian states reads
\begin{align}\label{eq:Gaussian_pure_a}
 \sI_\mathrm{pure} = \frac{1}{4} \tr\big((\ssp \dt\Gamma \Gamma^{-1})^2\big)+2\ssp \dt{\boldsymbol \delta}^T \Gamma^{-1} \dt{\boldsymbol \delta}\;,
\end{align}
which coincides with Eq.~(8) in~\cite{pinel_ultimate_2012}, or equivalently,
\begin{align}\label{eq:Gaussian_pure_b}
 \sI_\mathrm{pure} &= \frac{1}{4}  \tr\big((\ssp \dt\Gamma J)^2\big)+2\ssp \dt{\boldsymbol \delta}^T\! J\ssp \Gamma J^T \dt{\boldsymbol \delta}\;.
\end{align}
%For the degenerate case where $\Gamma_\std = S \Gamma S^T = \lambda I$, we have $\lambda\ssp \Gamma^{-1} = S^T\! S$, and consequently,
%\begin{align}
% \tr(\ssp \dt\Gamma_\std^2)  = \lambda^2 \tr\big((\ssp \dt\Gamma \Gamma^{-1})^2\big)\;.
%\end{align}
% Note that Eqs.~(\ref{eq:Gaussian_pure_a}) and (\ref{eq:Gaussian_pure_b}) work for all pure Gaussian states, not restricted to those driven by Gaussian unitaries.

\section{Conclusion}
\label{sec:conclusion}

For a quantum state in exponential form, we give expressions for the SLD, see Eqs.~(\ref{eq:expansion_L_a}) and (\ref{eq:expansion_L_b}), and the quantum Fisher information, see Eq.~(\ref{eq:Fisher_information}).  All these expressions are explicit and are useful for quantum-metrology problems with Gaussian or general thermal states (but are not restricted to these two kinds of states).  We give the quantum Fisher information, see Eq.~(\ref{eq:fisher_gaussian_generator}), for a Gaussian state in terms of its generator.  Using a different approach, we derive an equation for the SLD of an arbitrary Gaussian state in terms of its moments, confirming a recent result by Monras~\cite{monras_phase_2013}.  We find that the resulting equation is symplectic-covariant and can be solved exactly in the basis where the covariance matrix is in the standard form.  Furthermore, the Fisher information in terms of the moments of a general Gaussian state is calculated; special cases are discussed, which include pure, degenerate, and very noisy Gaussian states.

\begin{acknowledgments}
Special thanks go to C.~M. Caves for his very useful advice. The author also thanks J.~Combes, M.~D. Lang, and I.~H. Deutsch for helpful and enlightening discussions.  This work was supported by National Science Foundation Grant Nos.~PHY-1314763 and PHY-1212445 and Office of Naval Research Grant No.~N00014-11-1-0082.
\end{acknowledgments}

% \bibliographystyle{apsrev4-1_with_title}
% \bibliography{metro_exp}

%merlin.mbs apsrev4-1.bst 2010-07-25 4.21a (PWD, AO, DPC) hacked
%Control: key (0)
%Control: author (72) initials jnrlst
%Control: editor formatted (1) identically to author
%Control: production of article title (1) required
%Control: page (0) single
%Control: year (1) truncated
%Control: production of eprint (0) enabled
%

\end{document}